\documentstyle[12pt, epsf, fleqn]{article}
\evensidemargin=\oddsidemargin
\begin{document}

\Large 
{\bf Spectral Properties of a Hybrid Thermal-Nonthermal 

Isotropic Plasma}
\vspace{1.0cm}

\rm
\normalsize
D. Lin and E. P. Liang

\vspace{0.2cm}
\small
{\it Department of Space Physics and Astronomy, Rice University, Houston, TX 77005-1892, USA. 
\hspace{0.3cm}email: lin@spacsun.rice.edu. }

\normalsize
\vspace{0.5cm}

Received $\underline{\ \ \ \ \ \ \ \ \ \ \ \ \ \ \ \ \ \ \ \ \ \ }$; \hspace{0.5cm} Accepted 
$\underline{\ \ \ \ \ \ \ \ \ \ \ \ \ \ \ \ \ \ \ \ \ \ \ }$

\vspace{1.0cm}
\large
\parbox{0.8\hsize}{
{\bf ABSTRACT}

\vspace{0.2cm}
\normalsize
\parindent=0pt
The spectral properties of a hybrid steady thermal-nonthermal isotropic plasma are studied. The involved emission mechanisms are thermal bremsstrahlung, nonthermal bremsstrahlung, cyclotron and nonthermal synchrotron. Formulas are derived to calculate the emissivities and absorption coefficients of all these processes. Calculations show that the nonthermal synchrotron and the thermal bremsstrahlung dominate over most of the frequency range and that the cyclotron and the nonthermal bremsstrahlung are important only at the two ends of the spectrum. The combination of the two dominant processes could generate the synchrotron self-absorbed spectrum, the Rayleigh-Jeans spectrum and an absorbed flatter spectrum which is seen in some solar flare radio spectra. Absorbed spectra steeper than the three types of spectra are rare. The conditions for these three possible spectral types are given. Cyclotron harmonic features are seen in the emergent spectra from the hybrid plasmas. These harmonic features, useful in determining the magnetic field, exist only under certain conditions which are explored in this paper.

\vspace{0.3cm} \large
\parindent=0pt
{\bf Key words}: \normalsize radiation mechanisms: thermal -- radiation mechanism: nonthermal -- plasmas. 
}
\newpage
\normalsize
\parindent=12pt
\section{INTRODUCTION}

In a hybrid thermal-nonthermal steady-state plasma, thermal and nonthermal electrons coexist.
We Assume that the electron distribution is isotropic and can be described by
\begin{equation}
f( \gamma ) =\left \{ \begin{array}{ll}
                       f_{th}( \gamma )=A  \gamma  \sqrt{\gamma^{2}-1} e^{-\frac{( \gamma-1 )}{T_{e}}}& \gamma \leq \gamma_{th} \\ 
f_{nth}( \gamma )= B  \gamma^ {-p}     &            \gamma \geq \gamma_{th}
\end{array} \right .  
\end{equation}
where $\gamma$ is the Lorentz factor, $T_e$ is the temperature of thermal electrons in unit of $mc^2$, $f(\gamma)$ is normalized to 1, $A$ and $B$ are determined by the normalization and the continuity condition at
$\gamma_{th}$. This hybrid distribution may represent the particle distribution of many astronomical sources. From many RS CVn binary systems, good correlations of radio emission with x-ray and UV emission are observed (Drake, Simon \& Linsky 1989, 1992). Spectra show that the x-ray emission is due to thermal bremsstrahlung, and that the radio emission is produced by nonthermal power-law synchrotron emission (Drake et al. 1992). The correlations imply the co-existence of thermal and nonthermal electrons. In dense interstellar shocks,  thermal bremsstrahlung is thought to be the major emission mechanism (Neufeld \& Hollenbach 1996),  but the observed spectral index of -0.7 in radio bands (Curiel et al. 1993), which exceeds the limit of the thermal bremsstrahlung spectral index, implies that nonthermal electrons may also present in the plasma. In solar flares, the radio spectra and x-ray spectra indicate that the sources are hybrid thermal-nonthermal plasmas (Benka \& Hollman 1992, 1994). In some other astrophysical sources with a high Thomson depth such as x-ray sources (Crider et al. 1997) and  gamma ray bursters  (Tarvani 1996,  Liang 1997), the observed spectra suggest that the sources are hybrid thermal-nonthermal plasmas. 
Such  broad presence of the hybrid plasma makes it important for us to study its emission and absorption processes and the emergent self-absorbed spectrum
in the general case. 

The major emission processes in the hybrid plasma are gyrosynchrotron and bremsstrahlung. The gyrosynchrotron emission includes the cyclotron emission by the nonrelativistic electrons and the synchrotron emission by the relativistic electrons. The bremsstrahlung includes ion-electron bremsstrahlung, pair bremsstrahlung and electron-electron bremsstrahlung. But the electron-electron bremsstrahlung is weaker than the ion-electron bremsstrahlung at low energies (Haug 1985), which is of most concern in this paper for the comparison with gyrosynchrotron processes. Therefore, we will neglect the electron-electron bremsstrahlung.

Because of the complexity of calculating the emissivities of the two emission processes, we will confine our discussion to a certain physical regime which will simplify the calculation but still covers a wide range of astronomical phenomena. The magnetic field is assumed to be less than $10^{10} $ Gauss so that no quantum effects have to be considered. We will not consider the Compton scattering, so the Thomson depth $\tau_T \ll 1$. We will assume that the thermal electrons are nonrelativistic, or $kT_e \ll m_ec^2$.  The size of the source R is small enough that the photon escaping time $\frac{R}{c}$ is much less than the time scale of the electron distribution evolution. So the source can be treated as a steady-state one. The formulas we will review for the emissivities are applicable to arbitrary electron distribution functions. The nonthermal fraction $F_{nth} $, which is defined as the fraction of electrons with $\gamma > \gamma_{th}$, can be any value between 0 and 1. However, to be a meaningful hybrid source, it must includes enough thermal electrons, otherwise no thermal peak is seen in the particle distribution function. Therefore, we will constrain our discussions to $F_{nth} < 10\%$. We also assume that the source has a very small fraction of positrons.

The combination of gyrosynchrotron and bremsstrahlung emissions by the hybrid 
plasma can produce some distinct spectra which already have some useful applications (Benka \& Holman  1992, 1994). The gyrosynchrotron emissions by both thermal electrons and nonthermal electrons explain the harmonic features at the cyclotron frequencies (Benka \& Holman  1992), which are useful in determining the magnetic field of the sources. And also, the bremsstrahlung spectra by the hybrid electrons fit the soft and hard x-ray spectra of solar flares very well (Benka \& Holman 1994). However,  gyrosynchrotron or bremsstrahlung alone is not enough to explain another spectrum feature of solar flares called spectrum flattening (Lee, Garry \& Zirin 1994). Ramaty \& Petrosian (1972) has suggested that power-law synchrotron emissions absorbed by thermal bremsstrahlung can produce the flat spectra. They showed that the solar conditions favor such mechanism. However, the spectrum flattening should not be a unique property of solar flares. The flattened spectra should also be visible in 
other hybrid sources. The general conditions have to be found that favor such a flat spectrum. To find such conditions is equivalent to determine which, gyrosynchrotron or bremsstrahlung, dominates in the hybrid plasma. In this article, We will calculate the emissivities of gyrosynchrotron and bremsstrahlung emissions to determine the dominant processes and their conditions. We will also discuss the harmonic features and the conditions under which the harmonic features exist. 

In section 2, we will review the formulas for calculating the emissivities and the absorption coefficients of gyrosynchrotron and bremsstrahlung emissions. In section 3, we will show the general features of the emerging spectra from a hybrid plasma and  the dominant emission mechanisms. In section 3, we will find the conditions for the spectrum flattening in a hybrid plasma. In section 4, we will discuss the harmonic features.

\section{FORMULAS FOR EMISSIVITIES AND ABSORPTION COEFFICIENTS}
The emissivities $j_\nu$ and absorption coefficients $\alpha_\nu$ of gyrosynchrotron and bremsstrahlung emissions for an arbitrary normalized electron distribution 
function $f(\gamma)$ can be obtained from the following general expressions (Rybicki \& Lightman 1979)
 \begin{equation}
j_\nu = \frac{n_e}{4\pi} \int f(\gamma) P(\gamma, \nu) d\gamma 
\end{equation}
 \begin{equation}
 \alpha_\nu = \frac{n_e c^2}{8\pi h \nu^3} \int f^\prime (\gamma) P(\gamma, \nu) d\gamma  
\end{equation}
where $ f^\prime(\gamma) = \big( \frac{f(\gamma^*)}{\gamma^* \sqrt{{ \gamma^*}^2 - 1}} - \frac{f(\gamma)}{\gamma\sqrt{{ \gamma}^2 - 1}}\big) \gamma\sqrt{ \gamma^2 - 1} $ and $\gamma^* = \gamma - \frac{h\nu}{mc^2}$. $P(\gamma, \nu) $ is the power emitted by a particular electron.
The equations suggest that the formulas for absorption coefficients are easy to obtain once we have the  emissivity formulas. This property enables us to focus the  discussions in the following text on the calculations of emissivities. We will review the formulas for the emissivities of gyrosynchrotron and bremsstrahlung separately in the following two sub-sections.

The physical scenario that we discuss in this section and the following sections is a uniform and isotropic plasma cloud with a uniform magnetic field $B$ in it. The length of the line-of-sight inside the cloud is $L$. The spectrum emerging from the cloud is calculated from

\begin{equation}
 I_\nu = \frac{j_\nu}{\alpha_{\nu}}( 1 - e^{-\alpha_\nu L})
\end{equation}

\subsection{Gyrosynchrotron }

The emissivity for gyrosynchrotron emission can be calculated from the following equation which is derived in Brainerd (1985)
\begin{equation}
j_{\nu}^{gyr} = \frac{e^2\nu \pi n_e}{c} \sum_{n}^{\infty} \int_{\beta_{-}}^{\beta_{+}} \frac{\gamma^3 f(\gamma)}{\gamma_n \sqrt{\gamma^2 - 1}} \big[ ( \gamma_{||}^{-2} - \gamma^{-2} )J_{n}^{\prime 2}(\xi) + \frac{( cos\phi - \beta_{||})^2} {sin^2\phi}J_{n}^2(\xi)\big] d\beta_{||}  
\end{equation}
where $ n > \frac{\nu}{\nu_c} sin\phi  $, $ \gamma = \frac{\gamma_n}{1-\beta_{||} cos\phi} $, $ \gamma_n = \frac{n
\nu_c}{\nu}$, $\gamma_{||} = ( 1- \beta_{||}^2 )^{-\frac{1}{2}}$, $ \xi = \frac{\nu}{\nu_c}sin\phi\sqrt{ \gamma^2(1-\beta_{||}^2) - 1}$, $\nu_c = \frac{eB}{2mc\pi} $, and the integral limits are 

\begin{equation} 
\beta_{\pm} = \frac{cos\phi \pm \gamma_n \sqrt{\gamma_n^2 - sin^2\phi }}{ cos^2\phi + \gamma_n^2}  \end{equation}
The advantage of using $\beta_{||} $ as the integral variable instead of $\gamma $ (Ramaty 1969) is that no special treatment is necessary when $\phi = \frac{\pi}{2} $, where $\phi $ is the radiation angle relative to the magnetic field.

\subsection{Bremsstrahlung }

Let us first discuss ion-electron bremsstrahlung, then we will have a brief discussion of pair bremsstrahlung.

Due to the difficulty of evaluating the exact cross section for ion-electron bremsstrahlung, we use the Born approximation. But, in those cases that the electrons lose most of their energy, the Born approximation is no longer valid and the Coulomb correction is needed  (Jauch \& Rohrlich 1976). However, according to Jung (1994), the correction is only of the order of 10\% for nonrelativistic plasmas. So a low frequency approximation ($ h\nu \ll ( \gamma -1 )mc^2 $) will give satisfactory results. Under this assumption, the bremsstrahlung cross section for relativistic electrons is obtained from  Jauch \& Rohrlich (1976).
\begin{equation}   \frac{d\sigma}{d\epsilon} = \frac{16 \alpha z^2r_0^2 }{3 \epsilon \beta^2} \{ \frac{\beta^2}{4} - \frac{3}{4} + \frac{3}{4\gamma^2}ln(\gamma(1+\beta)) [ \frac{1}{\beta} - ln(\gamma(1+\beta)) ]+ ln\frac{2\gamma^2\beta^2mc^2}{\epsilon} \} 
\end{equation}
where $\alpha $ is the fine structure constant, $r_0$ is the electron classical radius, and $\epsilon = h\nu $. 
Equation (7) gives the correct limit of the cross sections in the two extreme limits of  $\beta \ll 1$ and $\gamma \gg 1$ (Zheleznyakov 1996).   

The emissivity of the ion-electron bremsstrahlung is given by

 \begin{equation} 
 j_\nu^{i+e} = \frac{\hbar n_e n_i c }{2} \int^{\infty}_{1+ \frac{\epsilon}{mc^2}} \beta f(\gamma) \epsilon \frac{d\sigma}{d\epsilon }d\gamma 
 \end{equation}

The pair bremsstrahlung is much more complicated for mildly relativistic leptons. But, according to Haug (1985), the emissivity of non-relativistic pair plasma has the same frequency dependence as that of ion-electron bremsstrahlung. The two differ only by factor of $2\sqrt{2} $. This enables us to treat the pair bremsstrahlung in a simple way. We assume that the cross section of pair bremsstrahlung is just a factor of $2\sqrt{2} $ higher than that of the ion-electron bremsstrahlung. Then we have the combined emissivity of ion-electron bremsstrahlung and pair bremsstrahlung 

\begin{equation}
j_\nu = \frac{ \hbar  n_e  ( n_i+ 2\sqrt{2} n_{e^+}) c }{2} \int^{\infty}_{1+ \frac{\epsilon}{mc^2}} \beta f(\gamma) \epsilon \frac{d\sigma}{d\epsilon }d\gamma 
\end{equation}
So including the pair bremsstrahlung or not won't qualitatively change the conclusions we are going to reach about the emission properties. In the following sections, we will set $  n_{e^+} $=0.

 \section{SPECTRAL SHAPES}
In order to explore the spectral features of  the hybrid plasma, emissivities and absorption coefficients are numerically calculated under various conditions based on the equations in section 2.

Two extreme test cases are first computed: a purely thermal source and a mainly power-law nonthermal source. For the thermal case, the temperature of the source is set to 10 keV. It can be treated as non-relativistic. Simple formula is available for calculating its bremsstrahlung emissivity (Rybicki and Lightman 1979). The results are shown in Fig.1. The overall emission is dominated by bremsstrahlung. Gyrosynchrotron emission dominates only in a small range of frequencies. The Rayleigh-Jeans spectrum is generated in the low frequency even though two different emission mechanisms are involved. This is what we expect for the optically thick thermal radiation. The source conditions and the calculation results of the mainly power-law case are shown in Fig. 2. The electron distribution in Fig. 2c is achieved by setting $T_e = $ 1000 keV and $\gamma_{th} $=1.2 in equation (1).  The bremsstrahlung emission is negligible in this case. The typical power-law synchrotron spectrum is seen in Fig. 2b. The harmonic features at the cyclotron frequency $\nu_c$ are due to the treatment of the low energy cut-off in the electron distribution (Fig. 2c).

Between the two extreme cases, various spectra may be generated by using different $B$, $n_e$ and $F_{nth} $. But generic shapes are shown in Fig. 3, where 99\% of the electrons are thermal and 1\% nonthermal. The solid line represents the spectrum from a thick source and the dash line for a thin source. The overall spectral shape is mainly determined by the two dominant processes: thermal bremsstrahlung and nonthermal synchrotron.  But the contributions of the nonthermal bremsstrahlung at the high energies ($h\nu > kT_e $) and thermal cyclotron near the cyclotron frequency $\nu_c $ are significant. In the low frequency range ($\nu > \nu_c $), the emission and absorption are dominated by the nonthermal synchrotron processes. The low frequency spectrum is a self-absorbed synchrotron spectrum. At higher frequencies, the thermal bremsstrahlung absorption would dominate over the synchrotron absorption, but the emissivity of the synchrotron is still larger than that of the bremsstrahlung. The nonthermal synchrotron emission absorbed by the thermal bremsstrahlung process is much flatter than the synchrotron self-absorption spectrum. At still higher frequencies, the thin source becomes optically thin, so the optically-thin synchrotron spectrum and then bremsstrahlung spectrum are seen; for the thick source, Rayleigh-Jeans spectrum is seen before the source finally turns optically thin. At the two ends of the spectra, the other two processes dominate, the nonthermal bremsstrahlung and the thermal cyclotron. Though more accurate calculation of the nonthermal bremsstrahlung should include the electron-electron bremsstrahlung, our results should give the right order of estimation. The cyclotron harmonic features, unexpected in the purely thermal or purely power-law cases, are produced because of the mixture of the thermal and nonthermal electrons in the cloud.

The flattening feature, a distinct spectral property of the hybrid plasma,  has been observed in solar flare spectra (Lee  et al 1994). Model fittings of the flat spectra will be presented in our future paper. We will explore the conditions  in the next section under which the flat spectrum will be seen. The harmonic features, useful in determining the magnetic field, exist only under certain conditions. We will discuss it in section 5.

\section{THE SPECTRUM FLATTENING}

As stated in the previous section, the dominant emission processes in the present physical regimes are thermal bremsstrahlung and nonthermal synchrotron. Since the cut-off frequencies of these two processes are $\nu_h =\frac{ kTe}{h} $ and $\nu_c$ respectively, the following discussion will be confined to the frequency range between  $\nu_c$ and $\nu_h$.  The combined source function of these two 
processes is
\begin{equation}
 s = \frac{ j^{syn} + j^{ff} }{\alpha^{syn} + \alpha^{ff}} 
 \end{equation}
where $j^{syn}$ and $j^{ff} $ are the emissivities of the nonthermal synchrotron and the thermal bremsstrahlung respectively, and $\alpha^{syn} $ and $\alpha^{ff}$ are the absorption coefficients. 

When $ j^{syn} \gg j^{ff}$ and $\alpha^{syn} \gg \alpha^{ff}$, the source function will be the nonthermal synchrotron source function $s^{syn} = \frac{ j^{syn}}{\alpha^{syn} } $, which is a power-law function with a power-law index of 2.5. When $ j^{syn} \ll j^{ff}$ and $\alpha^{syn} \ll \alpha^{ff}$, the source function will be that of the thermal bremsstrahlung $s^{ff} = \frac{ j^{ff}}{\alpha^{ff} } $, which is the Rayleigh-Jeans source function. There are two other cases that have to be discussed in details.
\begin{itemize}
\item { \it flattening} case: if  $ j^{syn} \gg j^{ff}$ and $\alpha^{syn} \ll \alpha^{ff}$,
   \begin{equation}
s = \frac{ j^{syn}} {\alpha^{ff} } \propto \nu^{2.5-p/2} 
 \end{equation}
Since p is usually larger than 1, the source function is flatter than that of both the synchrotron self-absorbed emission and the black body emission. 

\item {\it steepening} case:   if  $ j^{syn} \ll j^{ff}$ and $\alpha^{syn} \gg \alpha^{ff}$,
 \begin{equation}   
s = \frac{ j^{ff}} {\alpha^{syn} } \propto \nu^{2+p/2} 
 \end{equation}
Contrary to the  {\it flattening} case, the source function here is much steeper and the slope can
 be $ > 2.5$ when $ p > 1$.

\end{itemize}

However, these two cases do  not necessarily exist in the hybrid plasma. For thermal bremsstrahlung and power-law synchrotron, we have
 \begin{equation}
\frac{j^{syn}/ j^{ff}}{\alpha^{syn}/\alpha^{ff}} =\frac{ 298.5 ({\rm keV} )}{kT_e(p+1)}\sqrt{\frac{\nu}{\nu_c sin\phi}} 
\end{equation}
Under the condition that $\nu \gg \nu_c $, the inequality $ \frac{ j^{syn}}{j^{ff}} < \frac{\alpha^{syn}}{\alpha^{ff}} $ is valid only when $ kT_e(p+1) \gg {\rm 298.5 keV} $. This condition can not be satisfied in most of the cases that we are interested in, such as solar flares in which the temperature is about 10 keV.  So we assume that for all $\nu>\nu_c $
\begin{equation}
\frac{j^{syn}}{j^{ff}} > \frac{\alpha^{syn}}{\alpha^{ff}}
  \end{equation}
This means that the {\it steepening} case is rare.
 
Since  $j^{ff}$ is much flatter than $j^{syn} $, we will have $j^{syn}(\nu) < j^{ff}(\nu)$ and $\alpha^{syn}(\nu) < \alpha^{ff}(\nu)$ if $j^{syn}(\nu_c) <  j^{ff}(\nu_c) $.
Therefore, for those sources that satisfy the following condition, the thermal bremsstrahlung dominates over the whole frequency range ($\nu_c, \nu_h $)
\begin{equation}
\frac{n_e}{F_{nth}} > k_1 \frac{ B \cdot sin^{\frac{p+1}{2}}\phi} {ln(1.97\times 10^{11} \cdot T_e) - lnB } \end{equation}
where $ k_1 = 1.194 \times 10^{19} \cdot ( 1.731 )^p \cdot \frac{p-1}{p+1} T_e^{1/2} \gamma_{th}^{p-1} \Gamma(\frac{3p+19}{12}) \Gamma(\frac{3p-1}{12})  $,  $F_{nth} $ is the non-thermal fraction, and $T_e$ is in unit of keV. For p = 6, we have $k_1 =  4.418\times 10^{20} T_e^{1/2} \gamma_{th}^5 $.

Similarly, $ \alpha_{ff}(\nu_{h} ) < \alpha_{syn}(\nu_{h}) $ sets a condition under which the synchrotron emission dominates in the hybrid plasma. That is, the thermal bremsstrahlung emission is negligible when
\begin{equation}
\frac{n_e}{F_{nth}} < k_2 (B sin\phi)^{\frac{p+2}{2}} \end{equation}
where $k_2 = 4.987\times 10^{16} \cdot (5.898\times 10^6)^p T_e^{\frac{3-p}{2}} (p-1) \gamma_{th}^{p-1} \Gamma(\frac{3p+2}{12}) \Gamma(\frac{3p+22}{12})  $. 
If p = 6, we have $k_2 = 2.631\times 10^{-14} T_e^{-\frac{3}{2}}\gamma_{th}^5 $.

When the source conditions are between the two limits defined by equation (15) and (16), the  { \it flattening} case could happen at some frequencies between $\nu_c$ and $\nu_h $. If  $ \alpha_{ff}( \nu_0) = \alpha_{syn}(\nu_0) $ at a particular frequency $\nu_0$, the source function above $\nu_0$ changes to the flattening source function until $j_{syn}(\nu_1) = j_{ff}(\nu_1) $, where $\nu_1 > \nu_0 $. Therefore, for the spectrum flattening to happen between $\nu_0$ and $\nu_1$, the source condition must satisfy the following relations

\begin{equation}  \frac{n_e}{F_{nth}}  > k_3(\nu_0) (B sin\phi )^{\frac{p+2}{2}} \end{equation}
and
 \begin{equation} \frac{n_e}{F_{nth}} <  k_4(\nu_1) (B sin\phi)^{\frac{p+1}{2}} \end{equation}
where $k_3 = 4.119\times10^{16} \cdot ( 8.4\times10^{6} )^{p/2} \cdot (p-1) T_e^{\frac{1}{2}} \gamma_{th}^{p-1} \nu_0^{-p/2} ln^{-1}(\frac{5.517\times 10^{17}T_e}{\nu_0}) \Gamma(\frac{3p+2}{12}) \Gamma(\frac{3p+22}{12}) $ and $k_4= 7.133 \times 10^{15} ( 2.897 \times 10^3 )^p  \frac{p-1}{p+1} T_e^{\frac{1}{2}} \gamma_{th}^{p-1} \nu_1^{-(p-1)/2} ln^{-1}(\frac{5.517\times 10^{17}T_e}{\nu_1}) \Gamma(\frac{3p+19}{12}) \Gamma(\frac{3p-1}{12}) $.
For p = 6, $k_3 = 3.060\times10^{38} T_e^{\frac{3}{2} }\nu_0^{-3} \gamma_{th}^{5}  ln^{-1}(\frac{5.517\times10^{17}T_e}{\nu_0}) $ and $k_4 = 5.777\times10^{36} \cdot T_e^{\frac{1}{2} }\nu_1^{-2.5} \gamma_{th}^{5} $ $\cdot ln^{-1}(\frac{5.517\times10^{17}T_e}{\nu_1})$.

Fig. 4 is a summary of these results. The solid lines ${\rm L_{ff}}$ and ${\rm L_{syn}}$, which are defined by equation (15) and (16) respectively, define a region where the { \it flattening} case is possible. Above the line ${\rm L_{ff}}$, the thermal bremsstrahlung dominates, and the power-law synchrotron dominates in the region below the line  ${\rm L_{syn}}$. The dash  lines in Fig. 4a are the dividing lines where the source function starts to be flat at $\nu_0$. That is, at the frequency $\nu_0 $, we have $\alpha^{syn}(\nu_0) < \alpha^{ff}(\nu_0) $ above the lines  and $\alpha^{syn}(\nu_0) > \alpha^{ff}(\nu_0)$ below the lines. Similar plots are shown in Fig. 4b, where the dot-dash lines are the dividing lines for $j^{syn}$ and $j^{ff}$ at the frequency of $\nu_1 $.

Fig. 4 is useful in understanding the source conditions and the radiation processes. Three examples are showed. One is a neutron star with a surface lepton density of  
$10^{23}{\rm cm}^{-3}$ and a magnetic field of $10^{10}$ Gauss (Friendlander 1985). The figure suggests that the nonthermal fraction must be larger than
 $ 0.1\%$ for synchrotron emission to dominate in this case. Another example is the RS CVn binary system, whose magnetic loop has a plasma density of $10^7{\rm cm}^{-3}$ and a magnetic field of 5 Gauss (Franciosini \& Drago 1995). If the nonthermal fraction is about $ 0.1\%$,  the spectrum flattening may happen at a frequency of about $10^{10} $ Hz. The solar flare case is also shown in the figure.

\section{THE HARMONIC FEATURES}
The harmonic features seen in Fig. 3 are produced by  gyrosynchrotron emissions of the hybrid electrons. We divide the emissivity of the gyrosynchrotron $ j^{gyr} $ into $ j^{gyr}_{th} $ contributed from the thermal electrons and $j^{gyr}_{nth} $ from the nonthermal electrons. Similarly, we have the absorption coefficients $\alpha^{gyr}_{th} $ and $\alpha^{gyr}_{nth} $. The source function is expressed as
\begin{equation}
s = \frac{j^{gyr}_{th} + j^{gyr}_{nth} }{\alpha^{gyr}_{th} + \alpha^{gyr}_{nth}}
\end{equation}
At the frequencies $\nu \geq \nu_c $, $j^{gyr}_{th} $ has very sharp peaks while $j^{gry}_{nth} $ changes relatively smoothly (Fig. 1a and Fig. 2a). The same patterns are observed with $\alpha^{gry}_{th}$ and $\alpha^{gry}_{nth} $. Under certain conditions, the thermal processes dominate at some frequencies (the peaks of the thermal harmonic features) and the nonthermal processes dominate at others (the valleys of the thermal harmonic features). So the source function switches back and forth between  $\frac{\j^{gyr}_{th}}{\alpha^{gyr}_{th}} = k \nu^{2} $ and  $\frac{\j^{gyr}_{nth}}{\alpha^{gyr}_{nth}} = k^\prime \nu^{2.5}, $ where $k$ and $k^\prime $ are different constants,  producing the sawtooth pattern. This is why we see the harmonic feature in Fig. 3. However, one of the following two conditions would prevent the harmonic features from happening:

\begin{itemize}
\item If the thermal fraction is too high, $j^{gry}_{th} $ and $\alpha^{gry}_{th} $ are larger than $j^{gry}_{nth} $ and $\alpha^{gry}_{nth} $ respectively at all frequencies around $\nu_c$, the source function would simply be a power-law function with an index of 2.
\item If the nonthermal fraction is too high, the source function would also be a power-law function but with an index of 2.5.
\end{itemize}

These two conditions are defined by $T_e $, $F_{nth}$ and $p$ in the electron distribute function. But, due to the facts that $p$ varies in a relatively small range of values and that $j^{gry}_{nth}(\nu_c) $ does not sensitively depend on $p$ (equation (15)), $p$ is not a critical factor.   The magnetic field $B$ and the electron number density $n_e$ are not the determinant factors either, because  $j^{gry}_{th}(\nu_c)$ and $j^{gry}_{nth}(\nu_c) $ have the same dependencies on $B$ and $n_e$, and,  so do  $\alpha^{gry}_{th}(\nu_c)$ and  $\alpha^{gry}_{nth}(\nu_c)$ (equation (5)). By fixing $B$, $n_e$ and $p$, we calculate the spectra with various $T_e$ and $F_{nth}$. Fig. 5 show the region between ${\rm C_{syn}}$, ${\rm C_{ff}}$ and ${\rm C_{h}}$ where the harmonic features are possible. 
The low limit of $F_{nth} $ (line ${\rm C_{ff}}$) increase dramatically with $T_e$. This is because the depth of the thermal harmonic features decreases with the temperature $T_e$ so that more fraction of nonthermal electrons are necessary to prevent the thermal processes from dominating in all frequencies around $\nu_c$. The high limit of $F_{nth} $ (line ${\rm C_{syn}}$) decreases very little with $T_e$. This is due to the slow decrease of the thermal harmonic peak with $T_e$. When the temperature is too high ($T_e > 120 $ keV), the thermal harmonic peaks are almost smoothened out. No significant harmonic features are observed.  In many astrophysical scenarios like the solar flares, the electron temperature is about 10 keV. A very small fraction of nonthermal electrons in a thermal plasma are enough to produce the harmonic feature.

Constraints on $n_e$ and $B$ will come in when bremsstrahlung emissions and plasma effects are taken into account. Bremsstrahlung would dominate over nonthermal gyrosynchrotron when ($B$, $n_e$) is above ${\rm L_{ff}}$ in Fig. 4. So for the harmonic features to be possible,  ($B$, $n_e$) must be below 
 ${\rm L_{ff}}$. The horizontal line ${\rm C_h}$ in Fig. 5 is determined by  ${\rm L_{ff}}$ with $B$ = 400 Gauss and $n_e$ = $10^{10} {\rm cm}^{-3} $. Below the line, thermal gyrosynchrotron and thermal bremsstrahlung dominate, so the source function is just the Rayleigh-Jeans spectrum. The plasma effects require $\nu_c > \nu_{pe} $, where $\nu_{pe} $ is the plasma frequency. This requires $n_e < 9.898 \times 10^4 B^2 $, where $n_e$ and $B$ are in cgs units. This is a very strong constraint. We would not see the harmonic features in interstellar shocks because of the weak magnetic fields. The conditions of solar flares barely satisfy this condition.

The harmonic features may also be seen from other kinds of plasmas besides the hybrid thermal and nonthermal plasmas. A double-temperatured plasma 
is an example, in which the source function switches between the two black-body source functions with different temperatures. 
This may lead to the harmonic features.

 \section{ DISCUSSION }

In this paper, we review the formulas for calculating the emissivities and absorption coefficients of gyrosynchrotron and bremsstrahlung emissions to explore the spectral properties of the hybrid thermal-nonthermal plasma. The radiation processes in the hybrid plasma consist of thermal gyrosynchrotron, nonthermal synchrotron, thermal bremsstrahlung and nonthermal bremsstrahlung. Calculations showed that  nonthermal synchrotron and thermal bremsstrahlung dominate over most of the frequency range and that cyclotron and nonthermal bremsstrahlung are important only at  the two ends of the spectrum. At low frequencies the combination of the two dominant processes could generate the synchrotron self-absorbed spectrum, the Rayleigh-Jeans spectrum and a flatter spectrum which is seen in some solar flare radio spectra.  Spectra steeper than these three type of spectra are rare. The conditions for the three possible spectral types are given. Harmonic features, which is useful in determining the magnetic field, may be seen in the self-absorbed spectra of hybrid plasmas at the cyclotron frequency. To produce the harmonic features, the conditions must be satisfied that the magnetic field is strong enough to overcome the plasma effects, that the temperature must be less than 120 keV, and that
the nonthermal fraction $F_{nth}$ must be in the range shown in Fig. 5.

Throughout this paper we have assumed that the Thomson depth is negligible.
In some astronomical sources, however, the Thomson depth is high. The spectrum would be significantly distorted by the Compton scattering. But, in the lower energy end of the spectrum where absorption dominates over scattering, the spectral shapes we discussed in the previous sections would still be valid. Moreover, since the Compton scattering conserves the photon number, the total scattered photon flux is still controlled by the emission and absorption processes. Therefore the results we obtained in this paper is still useful as the photon sources for  high Thomson depth cases. The effects of the Compton scattering on the emission and absorption processes will be presented in future papers.

\parindent=0pt
\vspace{0.5cm}

{\bf ACKNOWLEDGMENT.} 

\vspace{0.2cm}
This work was partially supported by NASA grant NAG 5-3824. EPL thanks V. Petrosian for helpful discussions.

\newpage

{\bf REFERENCES}

\vspace{0.3cm}
Benka S. G.,  Holman, G. D., 1992, ApJ, 391, 854

Benka S. G.,  Holman G. D., 1994, ApJ, 435, 469

Brainerd J. J., 1985, PhD Thesis, Harvard University

Crider A., Liang E. P., Smith I. A., Lin D., Kusunose M., 1997, in Dermer C. D.,

\hspace{0.5cm} Strickman M. S., Kurfess J. D., eds, AIP Conference Proc. 410, The Fourth

\hspace{0.5cm} Compton Symposium, Williamsburge, VA, p. 868

Curiel S., Rodriguez L. F., Moran J. M., Canta J., 1993, ApJ, 415, 191

Drake S. A., Simon T., Linsky J. L., 1989, ApJS, 71, 905

Drake S. A., Simon T., Linsky J. L., 1992, ApJS, 82, 311

Franciosini E., Drago F. C., 1995, A\&A, 297, 535

Friedlander M. W., 1985, Astronomy, Prentice-Hall, New Jersey

Haug E., 1985, A\&A, 148, 386

Jauch J. M., Rohrlich F., 1976,  The Theory of Photons and Electrons, Springer-Verlag, 

\hspace{0.5cm} New York, p. 369

Jung Y.-D., 1994, Phys. Plasmas, 1, 785

Lee J. W., Garry D., Zirin H., 1994, Solar Physics, 152, 409

Liang E. P., 1997, ApJ, 491, L15

Neufeld D. A., Hollenbach D. J., 1996, ApJ, 471, L45

Ramaty R., 1969, ApJ, 158, 753

Ramaty R., Petrosian V., 1972, ApJ, 178, 241

Rybicki G., Lightman A., 1979, Radiative  Processes in Astrophysics, Wiley, New York

Tavani M., 1996, Phys. Rev. Lett., 176, 3476

Zheleznyakov V. V.,  1996, Radiation in Astrophysical Plasmas,  Kluwer, Netherland

\newpage

{\bf CAPTIONS }

\vspace{0.5cm}
Fig. 1. The emissivity and the spectrum of a purely thermal source. a: the emissivity. 
It is dominated by the thermal bremsstrahlung except for the frequencies near the cyclotron frequency $\nu_c$. The solid dots are the calculated values from the simple formula of the thermal bremsstrahlung emissivity with the gaunt factor. b:  the spectrum. The low energy part of the spectrum is the typical Rayleigh-Jeans spectrum. The high energy part is optically thin, and Its form is basically identical to that of the emissivity. The source conditions: $n_e = n_p = 10^{10} {\rm cm}^{-3}$, $n_{e^+} = 0 $, $T_e = 10 $ keV, $B$ = 10 Gauss, $\nu_c = 2.8 \times 10^7$ Hz, plasma frequency $\nu_{pe} = 8.98 \times 10^{8} $ Hz, $L = 10^9$ cm and $\phi = 90^o$. 

\vspace{0.5cm}
Fig. 2. The emissivity and the spectrum of a power-law source. a: the emissivity. The gyrosynchrotron emission dominates the emissivity. The bremsstrahlung emission is negligible. The solid dots are the calculated values from the formula of equation (6.36) in Rybicki \& Lightman (1979) . b: the spectrum. The spectrum above the cyclotron frequency ($\nu > \nu_c)$ is the typical power-law synchrotron spectrum.  c: the electron distribution function used in this calculation. The source conditions: $n_e = n_p = 10^{10} {\rm cm}^{-3}$, $n_{e^+} = 0$, $B$ = 10 Gauss, $p$ = 3.5, $\nu_c = 2.8 \times 10^7$ Hz, $\nu_{pe} = 8.98 \times 10^{8} $ Hz,  $L = 10^9 $cm, and $\phi = 90^o$. 

\vspace{0.5cm}
Fig. 3. Typical spectral shapes from a hybrid thermal-nonthermal source. The spectrum can be divided into several parts each of which has different emission and absorption mechanisms as labeled in the figure. The solid line is for a thick source with $L =  10^{15}$ cm. The dash line represents a thin source with $L =  10^{9}$ cm. Other source conditions: $n_e = n_p = 10^{12} {\rm cm}^{-3}$, $n_{e^+} = 0$, $F_{nth} = 1\%$, $B$ = 1 Gauss, $p$ = 3.5, $\nu_c = 2.8 \times 10^6$ Hz, $\nu_{pe} = 8.98 \times 10^{9} $ Hz, $T_e$ = 10 keV, and $\phi = 90^o$.

\vspace{0.5cm}
Fig. 4. The dominating regions of the two dominant emission processes in $n_e$ - $B$ space. Below the solid line ${\rm L_{syn}}$, synchrotron dominates over the frequency range of ($\nu_c, \nu_h$). Above  the solid line ${\rm L_{ff}}$, thermal bremsstrahlung dominates. In the region between ${\rm L_{syn}}$ and ${\rm L_{ff}}$, either synchrotron or thermal bremsstrahlung dominates over a certain frequency range. In the upper panel a, for a certain frequency $\nu_0$, above the dash line $\alpha^{ff}(\nu_0) > \alpha^{syn}(\nu_0)$ and below the  dash line $\alpha^{ff}(\nu_0) <  \alpha^{syn}(\nu_0)$. In the lower panel b, similar dash-dot lines are drawn for $j^{ff}(\nu_1) = j^{syn}(\nu_1) $. The spectrum flattening may happen only in the region between  ${\rm L_{syn}}$ and ${\rm L_{ff}}$. The fixed source conditions are:   $T_e$ = 10 keV, $p$ = 6,  $\gamma_{th} = 1.18 $,  and $\phi = 90^o$. 

\vspace{0.5cm}
Fig. 5. The possible region for the harmonic features in $T_e $-$F_{nth}$ plane. Above ${\rm C_{syn}}$, the spectrum is the synchrotron self-absorbed spectrum. Below  ${\rm C_{ff}}$ and ${\rm C_{h}} $ are the Rayleigh-Jeans spectrum zone. ${\rm C_{h}} $ is determined by ${\rm L_{ff}}$ in Fig. 4.
Other source conditions are: $B$ = 400 Gauss, $n_e$ = $10^{10} {\rm cm}^{-3} $, $p$ = 6 and $\phi = 90^o$.

\newpage

\begin{figure}[h]
\epsffile{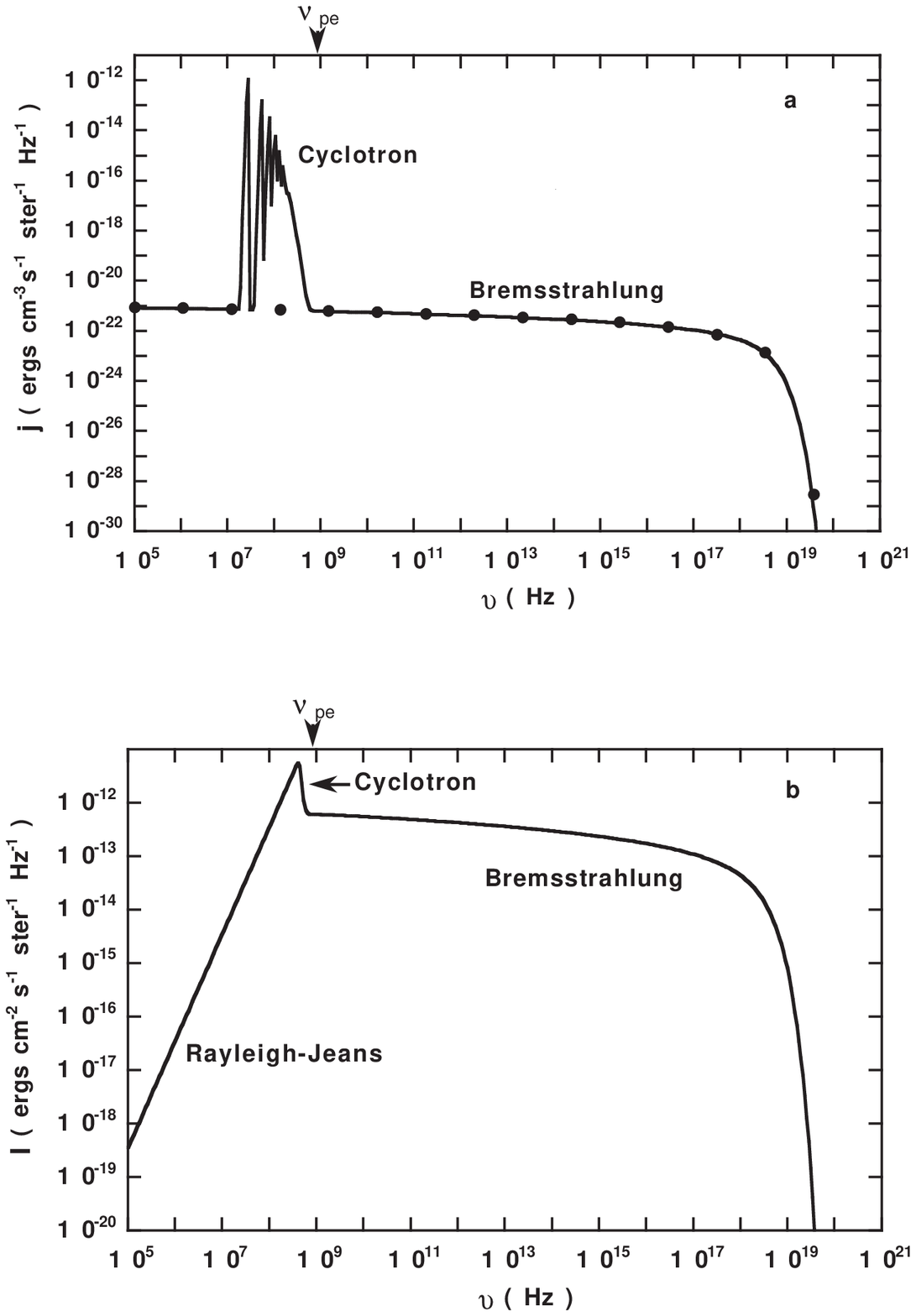}
\caption{}
\end{figure}

\newpage

\begin{figure}[h]
\epsffile{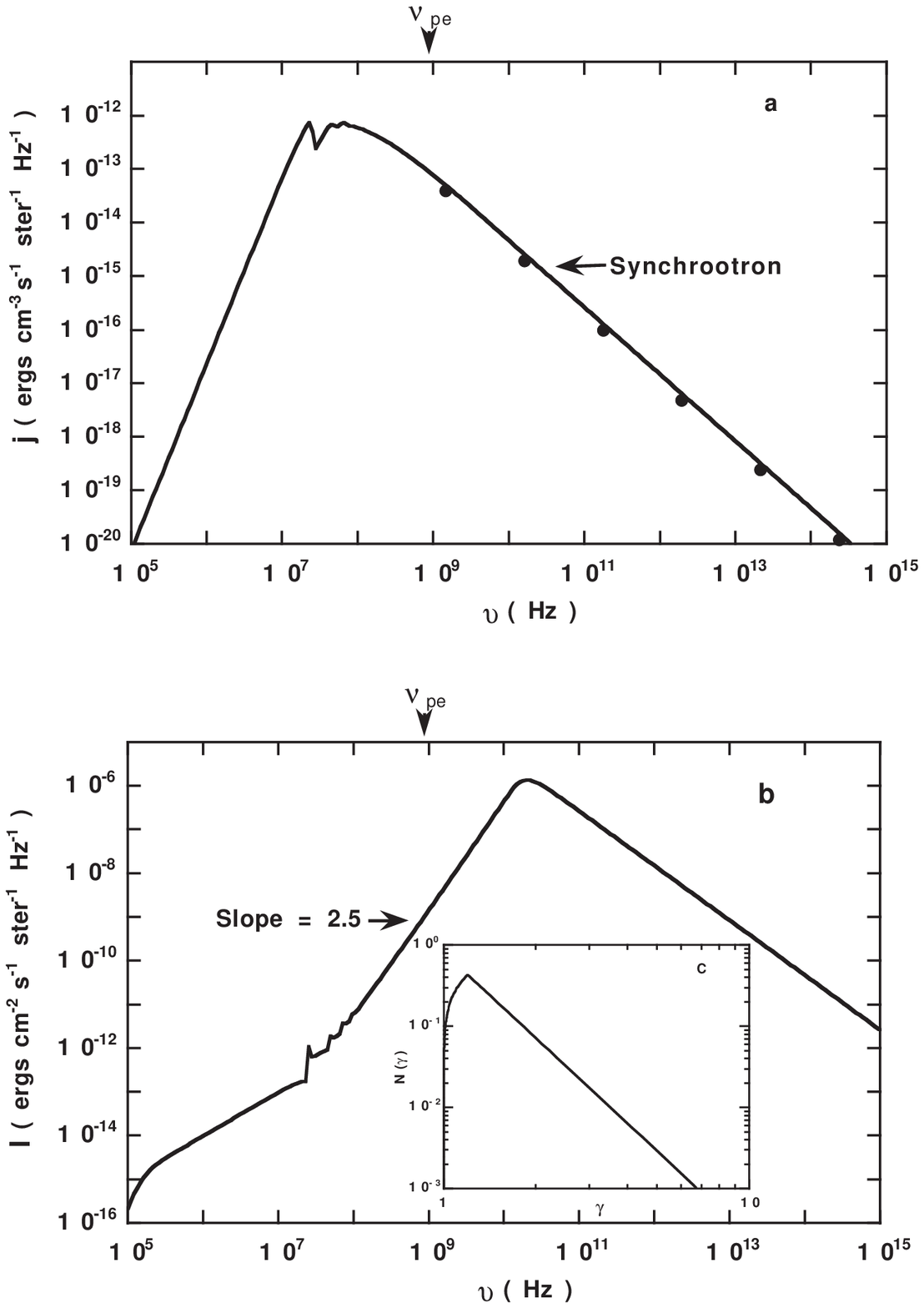}
\caption{}
\end{figure}
\newpage

\begin{figure}[h]
\epsffile{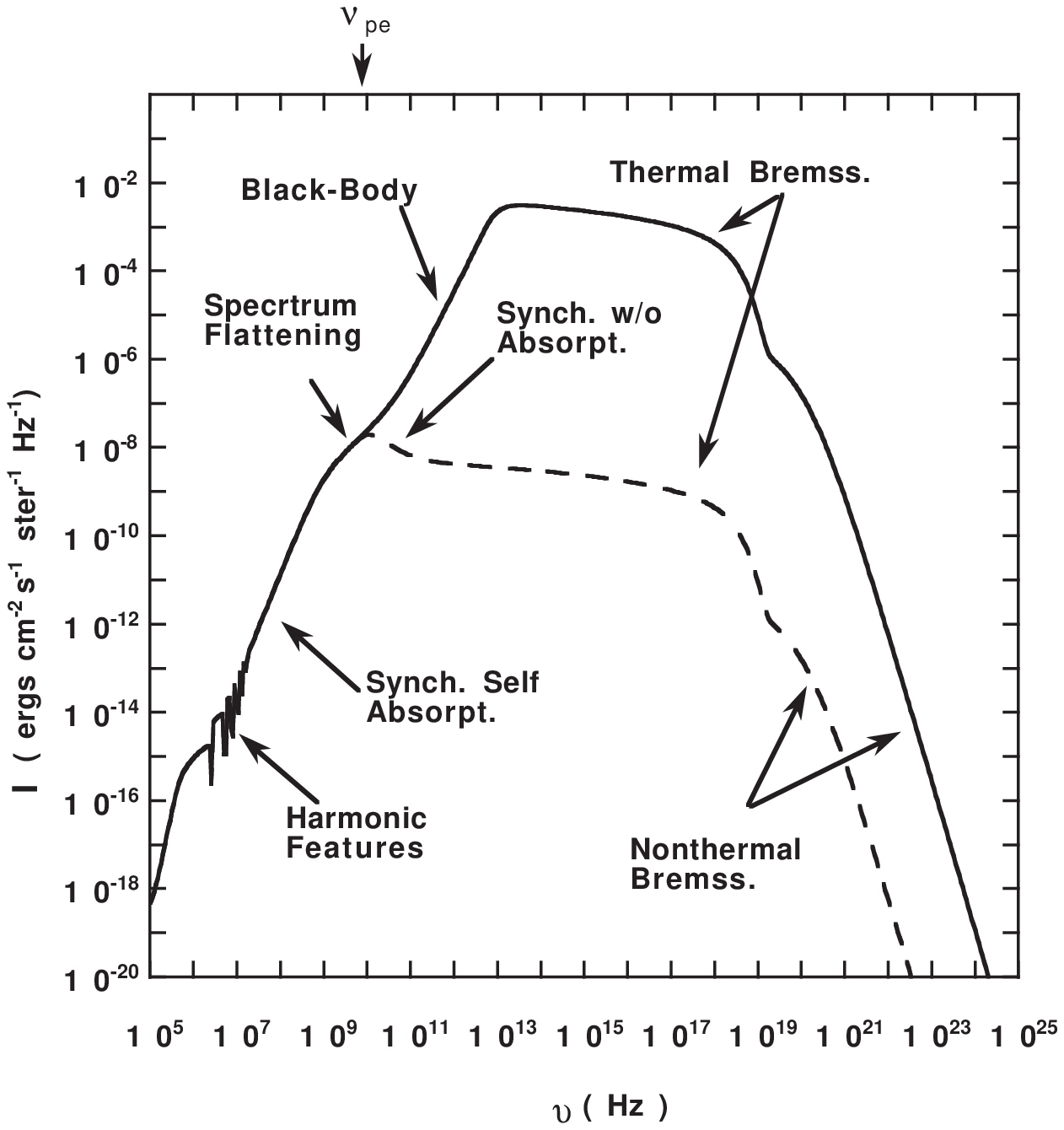}
\caption{}
\end{figure}
\newpage

\begin{figure}[h]
\epsffile{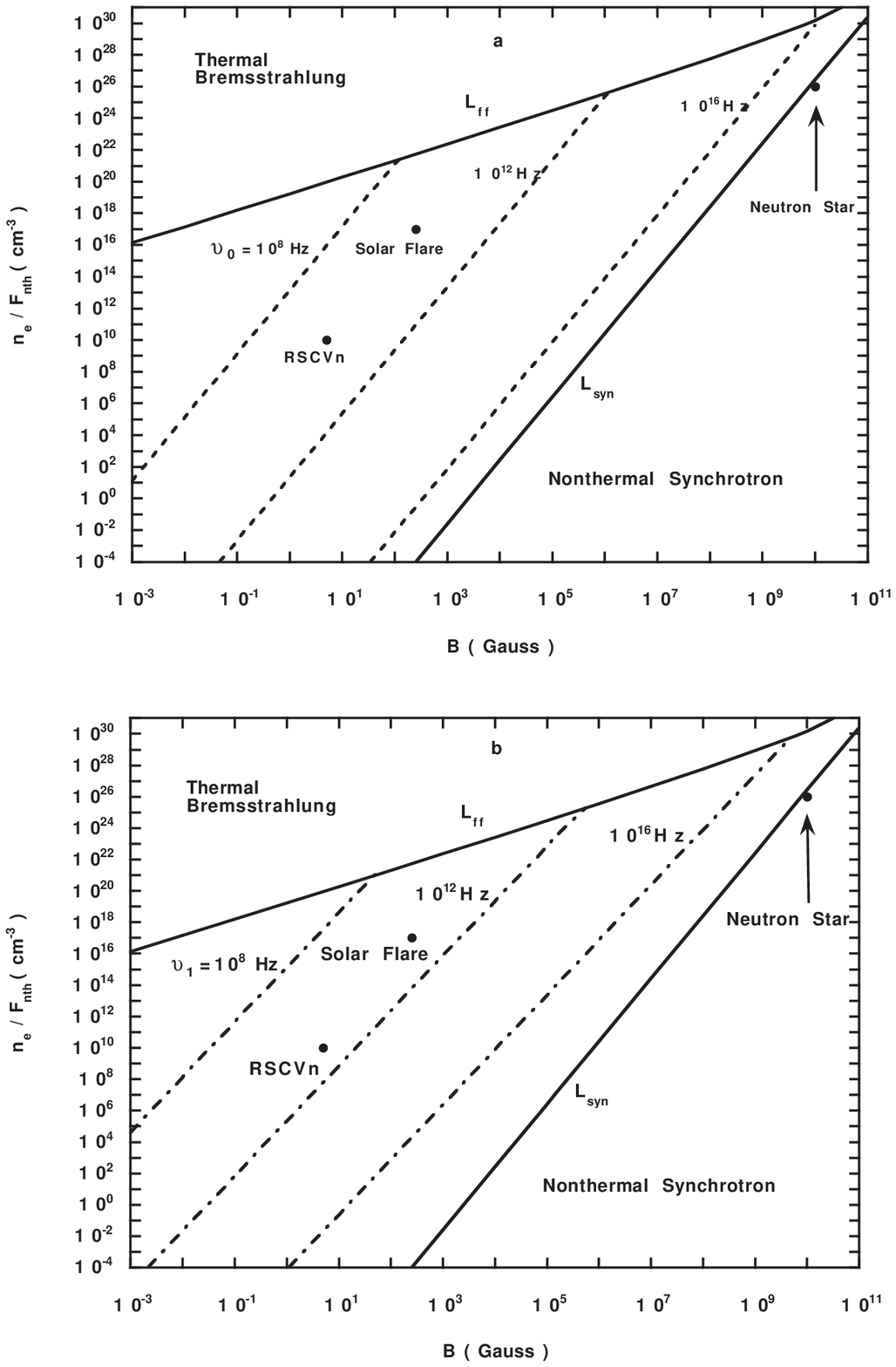}
\caption{}
\end{figure}
\newpage

\begin{figure}[h]
\epsffile{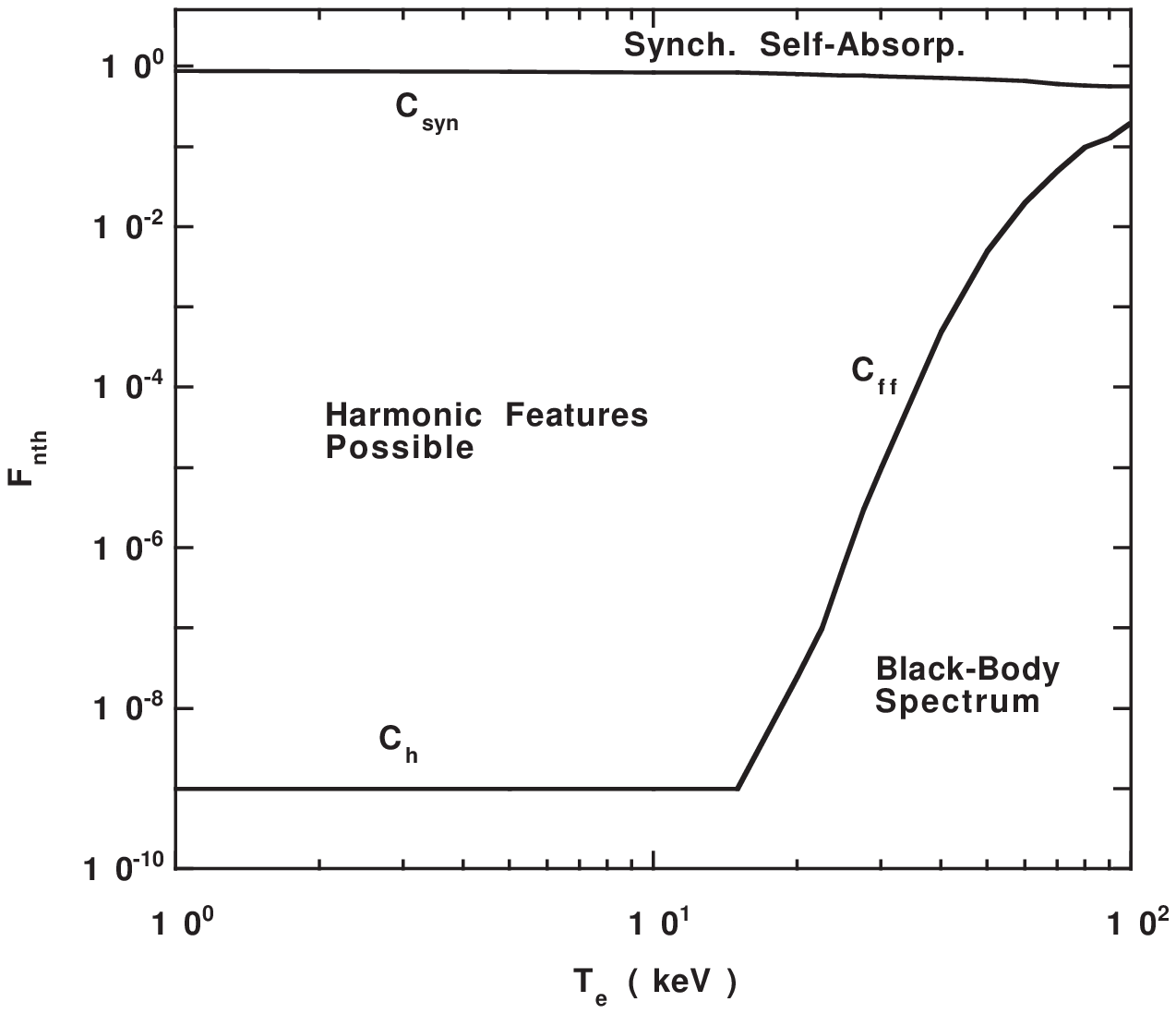}
\caption{}
\end{figure}

\end{document}